\documentclass[preprint]{elsarticle}
\journal{Physics Letters B}

\usepackage[english]{babel}
\usepackage{graphicx}
\usepackage{amsmath,amssymb,amsfonts,amstext}
\usepackage{hyperref}

\begin{document}
\begin{frontmatter}

\title{Regge-plus-resonance predictions for kaon photoproduction from the neutron}

\author{P.~Vancraeyveld\corref{author}}\ead{Pieter.Vancraeyveld@UGent.be}
\author{L.~De~Cruz}
\author{J.~Ryckebusch}
\author{T.~Van~Cauteren}
\address{Department of Physics and Astronomy, Ghent University, Proeftuinstraat 86, B-9000 Gent, Belgium}
\cortext[author]{Corresponding author}

\begin{abstract}
We present predictions for $n(\gamma,K^+)\Sigma^-$ differential cross sections and photon-beam asymmetries and compare them to recent LEPS data. We adapt a Regge-plus-resonance~(RPR) model developed to describe photoinduced and electroinduced kaon production off protons. The non-resonant contributions to the amplitude are modelled in terms of $K^+(494)$ and $K^{\ast+}(892)$ Regge-trajectory exchange. This amplitude is supplemented with a selection of $s$-channel resonance diagrams. The three Regge-model parameters of the $n(\gamma,K^+)\Sigma^-$ amplitude are derived from the ones fitted to proton data through $SU(2)$ isospin considerations. A fair description of the $n(\gamma,K^+)\Sigma^-$ data is realized, which demonstrates the Regge model's robustness and predictive power. Conversion of the resonances' couplings from the proton to the neutron is more challenging, as it requires knowledge of the photocoupling helicity amplitudes. We illustrate how the uncertainties of the helicity amplitudes propagate and heavily restrain the predictive power of the RPR and isobar models for kaon production off neutron targets.
\end{abstract}

\end{frontmatter}

Mapping out the baryonic spectrum remains a paramount issue in hadron physics. The masses, widths and transition form factors of the nucleon's excited states are invaluable input to models aimed at understanding the internal structure of baryons. In this regard, electromagnetic~(EM) kaon production plays a key role in the ongoing theoretical and experimental efforts to explore the dynamics of QCD in the confinement regime.

Electron accelerator facilities, such as ELSA, Jefferson Lab, MAMI and SPring-8, are making contributions to a ``complete'' kaon production experiment. Along with the unpolarised differential cross section, this requires the measurement of seven carefully chosen single and double polarisation observables~\cite{CompleteBarker,CompleteTabakin}. Ideally, this leads to an unambiguous determination of the reaction amplitude and, as such, stringent constraints on dynamical models. Thus far, the lion's share of research efforts has been directed towards reactions off proton targets. The complementary reaction on neutrons yields additional constraints that help to further pin down the underlying reaction dynamics. Moreover, the neutron channels are a crucial ingredient in the description of hypernuclear spectroscopy and quasi-free kaon production on nuclei. These reactions provide, amongst other things, access to the hyperon-nucleon interaction~\cite{HypernuclearReview}.

The presence of open strangeness in the final state of electromagnetic kaon production holds out the prospect of finding some elusive resonant states. Despite the publication of a large body of high-quality $p(\gamma^{(\ast)},K)Y$ data in recent years, phenomenological analyses have not led to an unequivocal outcome. Disentangling the relevant resonant contributions is challenging, because of the large number of competing resonances above the kaon production threshold. Moreover, the smooth energy dependence of the measured observables hints at a dominant role for the background, i.e.\ non-resonant, processes. Hence, the treatment of the background is pivotal for any model. In traditional isobar approaches, these non-resonant terms diverge as energy increases~\cite{StijnHyperonRes}. Over the years, several mechanisms to remedy this unrealistic behaviour have been proposed. The extracted resonance couplings, however, heavily depend on the background model~\cite{Stijn3bckgrModels,DaveGeneticAlg}.

At sufficiently high energies, the isobar description is no longer optimal. In this energy region, the kaon production amplitude can be elegantly described within the Regge framework, characterized by the exchange of whole families of particles, instead of individual hadrons~\cite{GuidalPhotoProdKandPi}. Interestingly, the Regge model, a high-energy theory by construction, allows to describe the gross features of the data in the resonance region~\cite{GuidalUpdate,RPRlambda,RPRsigma}. Extrapolating the Regge model to intermediate energies results in a reliable account of the kaon-production background within the resonance region and imposes a proper high-energy behaviour.

Building upon the work of Guidal et al.~\cite{PHDguidal,GuidalPhotoProdKandPi}, we model the $p(\gamma,K^+)\Sigma^0$ background amplitude by means of $K^+(494)$ and $K^{\ast+}(892)$ Regge-trajectory exchange in the $t$-channel~\cite{RPRsigma}. A gauge invariant amplitude is obtained by adding the electric part of the nucleon $s$-channel Born diagram. The strong forward-peaked character of the differential cross section provides powerful support for this approach. The exchange of a linear kaon Regge-trajectory 
\begin{equation}\label{eq:trajectory}
 \alpha_{K^{(\ast)+}}(t) = \alpha_{K^{(\ast)+},0} + \alpha'_{K^{(\ast)+}} \left( t - m^2_{K^{(\ast)+}} \right)\, ,
\end{equation}
with $m_{K^{(\ast)+}}$ and $\alpha_{K^{(\ast)+},0}$ the kaon's mass and spin, is realised through a Reggeized amplitude that combines elements of the Regge formalism and a tree-level effective-Lagrangian model. Reggeization amounts to replacing the standard Feynman ${(t-m^2_{K^{(\ast)+}})^{-1}}$ propagator by the corresponding Regge propagator
\begin{equation}\begin{split} \label{eq: reggeprop}
\mathcal{P}^{K^+(494)}_{\text{Regge}}(s,t) &= \left(\dfrac{s}{s_0}\right)^{\alpha_{K^+}(t)}
\dfrac{e^{-i\pi\alpha_{K^+}(t)}}{\sin\bigl(\pi\alpha_{K^+}(t)\bigr)} \; \dfrac{\pi \alpha'_{K^+}}{\Gamma\bigl(1+\alpha_{K^+}(t)\bigr)}  \,,
\\
\mathcal{P}^{K^{\ast +}(892)}_{\text{Regge}}(s,t) &= \left(\dfrac{s}{s_0}\right)^{\alpha_{K^{\ast+}}(t)-1} 
\dfrac{1}{\sin\bigl(\pi\alpha_{K^{\ast+}}(t)\bigr)} \; \dfrac{\pi \alpha'_{K^{\ast+}}}{\Gamma\bigl(\alpha_{K^{\ast+}}(t)\bigr)} \,,
\end{split}\end{equation}
with $s_0=1\,\text{GeV}^2$, $\alpha_{K^+}(t) = 0.70 \ (t-m_{K^+}^2)$ and $\alpha_{K^{\ast+}}(t) = 1 + 0.85 \ (t-m_{K^{\ast+}}^2)$, when $t$ and $m_{K^{(\ast)+}}^2$ are expressed in units of $\text{GeV}^2$. The data~\cite{Boyarski,Quinn} indicate that the trajectories are strongly degenerate. Consequently, the Regge propagators have either a constant or rotating phase. These phases cannot be deduced from first principles. In Ref.~\cite{RPRsigma}, we found the $p(\gamma,K^+)\Sigma^0$ data to be compatible with a rotating and a constant phase for the $K^+(494)$ and $K^{\ast+}(892)$ trajectories respectively. In our implementation of the Regge model, the operatorial structure of the amplitudes is dictated by an effective Lagrangian approach\footnote{Our choice of strong and electromagnetic interaction Lagrangians can be found in Ref.~\cite{RPRlambda}}, in which the $t$-channel propagators are replaced by the corresponding Regge ones. As a consequence, the amplitude corresponding to $K^{(\ast)+}$ exchange in the $t$-channel effectively incorporates the transfer of an entire trajectory. When considering the exchange of $K^+(494)$ and $K^{\ast+}(892)$ trajectories, the Regge model for $p(\gamma,K^+)\Sigma^0$ has a mere three parameters
\begin{equation}\label{eq:reggeparameters}
 g_{K^+\Sigma^0p}\, , \quad G^{v,t}_{K^{\ast+}\Sigma^0p} = \kappa_{K^{\ast+}K^+} \dfrac{e\,g^{v,t}_{K^{\ast+}\Sigma^0p}}{4\pi} \, ,
\end{equation}
with $g_{K^+\Sigma^0p}$, $g^v_{K^{\ast+}\Sigma^0p}$ and $g^t_{K^{\ast+}\Sigma^0p}$ the coupling constants at the strong interaction vertex and $\kappa_{K^{\ast+}K^+}$ the $K^{\ast+}(892)$'s transition magnetic moment.

The Regge model's amplitude can be interpreted as the asymptotic form of the full amplitude for large $s$ and small $|t|$. Owing to the $t$-channel dominance and the absence of a prevailing resonance, the Reggeized background can account for the gross features of the kaon production data within the resonance region~\cite{GuidalUpdate,RPRsigma}. Near threshold, the energy dependence of the measured differential cross sections exhibits structure which hints at the presence of resonances. These are incorporated by supplementing the background with a number of resonant $s$-channel diagrams. This approach was coined Regge-plus-resonance~(RPR) and has previously been applied to double-pion production~\cite{PHDholvoet}, as well as the production of $\eta$ and $\eta'$ mesons~\cite{RPReta}. We describe the resonant contributions using standard tree-level Feynman diagrams. By substituting $s-m_R^2 \rightarrow s-m_R^2+im_R\Gamma_R$ in the propagator's denominator, we take into account the finite lifetime of resonances with mass $m_R$ and width $\Gamma_R$. To limit the number of fit parameters, we keep the resonances' mass and width fixed at the values given in the Particle Data Group's Review of Particle Physics~(RPP)~\cite{pdg}. Each spin-$1/2$ resonance introduces one free parameter
\begin{equation}
 G_{N^{\ast}} = \kappa_{N^{\ast}p} \times g_{K^+\Sigma^0N^{\ast}}\, ,
\end{equation}
the product of the coupling constants at the electromagnetic and the strong interaction vertex. Spin-$3/2$ resonances have an additional degree of freedom at the photon vertex and give rise to two free parameters
\begin{equation}\begin{split}
 G^{(1)}_{N^{\ast}} &= \kappa^{(1)}_{N^{\ast}p} \times g_{K^+\Sigma^0N^{\ast}}\, ,\\
 G^{(2)}_{N^{\ast}} &= \kappa^{(2)}_{N^{\ast}p} \times g_{K^+\Sigma^0N^{\ast}}\, .
\end{split}\end{equation}
The most general interaction Lagrangian for spin-$3/2$ fields allows for an additional three degrees-of-freedom, often called \textit{off-shell} parameters, in the strong and EM vertices~\cite{Spin32lagrangian}. To ensure that the effects of the resonant diagrams fade at higher energies, we introduce a Gaussian form factor with a cutoff $\Lambda_{\text{strong}}\approx1.6\,\text{GeV}$ at the strong interaction vertices ~\cite{RPRsigma}. 

The dynamics of electromagnetic kaon production can be fairly involved, with several contributing nucleon and delta resonances that interfere with an eminent background. Disentangling these contributions is challenging. In the RPR approach, we seek to determine the resonant and non-resonant terms separately~\cite{RPRlambda,RPRsigma}. At sufficiently high energies ($\omega_{\text{lab}}\gtrsim4\,\text{GeV}$), a limited amount of 57 $p(\gamma,K^+)\Sigma^0$ data points are available, consisting of differential cross sections~\cite{Boyarski} and photon-beam asymmetries~\cite{Quinn}. These data show no resonant features and are used to constrain the three parameters of the Regge model in Eq.~\eqref{eq:reggeparameters}. In the resonance region, a large body of data is available~\cite{LEPSpho,CLASdcs06,CLASrec,GRAALrecpho}, against which we fit the resonance parameters, while keeping the background unaltered. In Ref.~\cite{RPRsigma}, we established the phases of the leading kaon trajectories. With the available 57 data points, it turned out impossible to single out a unique parametrization of the Regge model, as the sign of $G^{t}_{K^{\ast+}\Sigma^0p}$ remained undetermined. The two model variants, that yield an equally good description of the high-energy data, were labelled Regge-3 and Regge-4. Subsequently, we added resonances to the Reggeized background amplitude, identifying the $S_{11}(1650)$, $D_{33}(1700)$, $P_{11}(1710)$, $P_{13}(1720)$, $P_{13}(1900)$, $S_{31}(1900)$, $P_{31}(1910)$ and $P_{33}(1920)$ as essential contributions. These are established resonances with a 3- or 4-star status in the RPP~\cite{pdg}, except for the $P_{13}(1900)$ and $S_{31}(1900)$, which are 2-star resonances. Both the RPR-3 and RPR-4 models reach a goodness-of-fit of $\chi^2/d.o.f.=2.0$. We found no direct need to include ``missing'' resonances in the $K\Sigma$ channel. 

In order to assess the predictive power of the RPR model, we extended our formalism to kaon electroproduction in Ref.~\cite{RPRelectro}. The $Q^2$-dependence of the EM coupling constants was incorporated using transition form factors as computed in the Bonn constituent-quark model~\cite{BonnEMff}. Without refitting any parameters, we found that the RPR model gives a decent account of the available kaon electroproduction data. Kaon production off neutrons offers another opportunity to test the robustness of the RPR approach. In what follows, we will point out how the fitted RPR amplitude for the $p(\gamma,K^+)\Sigma^0$ channel can be transformed with an eye to predicting the $n(\gamma,K^+)\Sigma^-$ reaction. 

In order to relate $n(\gamma,K^+)\Sigma^-$ to $p(\gamma,K^+)\Sigma^0$, it suffices to convert the coupling constants which feature in the interaction Lagrangians. In the strong interaction vertex, we assume isospin symmetry to be exact. The hadronic couplings are proportional to the Clebsch-Gordan coefficients:
\begin{equation}\begin{split}
 g_{K\Sigma N^{(\ast)}} &\sim \left< I_K=\frac{1}{2}, M^I_K\,;\,I_{\Sigma} = 1, M^I_{\Sigma}\;\left|\;I_{N^{(\ast)}}=\frac{1}{2},M^I_{N^{(\ast)}}\right.\right>\, ,\\
 g_{K\Sigma \Delta^{\ast}} &\sim \left< I_K=\frac{1}{2}, M^I_K\,;\,I_{\Sigma} = 1, M^I_{\Sigma}\;\left|\;I_{\Delta}=\frac{3}{2},M^I_{\Delta}\right.\right>\, .
\end{split}\end{equation}
We adopt the following conventions for the isospin states of the $N^{(\ast)}$, $\Delta^{\ast}$, $K^{(\ast)}$ and $\Sigma$ particles,
\begin{equation}
 \begin{array}{rcr@{}l}
  p,K^{(\ast)+},N^{\ast+} &\rightarrow & &\left|I=\frac{1}{2},\right. \,M^I=+\left.\frac{1}{2}\right>\, ,\\\noalign{\medskip}
  n,K^{(\ast)0},N^{\ast0} &\rightarrow & &\left|I=\frac{1}{2},\right. \,M^I=-\left.\frac{1}{2}\right>\, ,\\\noalign{\medskip}
  \Lambda &\rightarrow & &\left|I=0,\right. \,M^I=0\left.\right>\, ,\\\noalign{\medskip}
  \Sigma^+ &\rightarrow &- &\left|I=1,\right. \,M^I=\left.1\right>\, ,\\\noalign{\medskip}
  \Sigma^0 &\rightarrow & &\left|I=1,\right. \,M^I=\left.0\right>\, ,\\\noalign{\medskip}
  \Sigma^- &\rightarrow & &\left|I=1,\right. \,M^I=-\left.1\right>\, ,\\\noalign{\medskip}
  \Delta^{\ast+} &\rightarrow & &\left|I=\frac{3}{2},\right. \,M^I=+\left.\frac{1}{2}\right>\, ,\\\noalign{\medskip}
  \Delta^{\ast0} &\rightarrow & &\left|I=\frac{3}{2},\right. \,M^I=-\left.\frac{1}{2}\right>\, .
 \end{array}
\end{equation}
The phase of the $\Sigma^-$ state is taken to be positive. With this choice, the Condon-Shortley phase convention dictates a minus sign for the $\Sigma^+$ state. The above leads to the following relations:
\begin{equation}\begin{split}\textstyle
 g_{K^{(\ast)+}\Sigma^-n} &= \sqrt{2}\,g_{K^{(\ast)+}\Sigma^0p}\, ,\\
 g_{K^{(\ast)+}\Sigma^-N^{\ast0}} &= \sqrt{2}\,g_{K^{(\ast)+}\Sigma^0N^{\ast+}}\, ,\\
 g_{K^{(\ast)+}\Sigma^-\Delta^{\ast0}} &= \dfrac{1}{\sqrt{2}}g_{K^{(\ast)+}\Sigma^0\Delta^{\ast+}}\, .
\end{split}\end{equation}

In contrast to the hadronic parameters, the relations between electromagnetic couplings have to be distilled from experimental information. The $N^{\ast}$ and $\Delta^{\ast}$ transition moments can be related to the photocoupling helicity amplitudes $\mathcal{A}^N_{J}$. One has
\begin{equation}\label{eq:helamp12}
 \mathcal{A}^N_{1/2} = \mp \dfrac{e}{2m_N}\sqrt{\dfrac{m_{N^{\ast}}^2-m_N^2}{2m_N}}\kappa_{{N^{\ast}}N}\, ,
\end{equation}
for spin-$1/2$ resonances and
\begin{equation}\begin{split}\label{eq:helamp32}
 \mathcal{A}^N_{1/2} &= \dfrac{e}{4m_{N^{\ast}}} \sqrt{\dfrac{m_{N^{\ast}}^2-m_N^2}{3m_N}} \left( \pm \kappa^{(1)}_{{N^{\ast}}N} - \dfrac{m_{N^{\ast}}(m_{N^{\ast}}\mp m_N)}{4m_N^2} \kappa^{(2)}_{{N^{\ast}}N}\right)\, ,\\
 \mathcal{A}^N_{3/2} &= \dfrac{e}{4m_N} \sqrt{\dfrac{m_{N^{\ast}}^2-m_N^2}{m_N}}  \left( \pm \kappa^{(1)}_{{N^{\ast}}N} \mp \dfrac{m_{N^{\ast}}\mp m_N}{4m_N} \kappa^{(2)}_{{N^{\ast}}N}\right)\, ,
\end{split}\end{equation}
for spin-$3/2$ resonances. In equations~\eqref{eq:helamp12} and~\eqref{eq:helamp32} the upper (lower) sign corresponds to positive- (negative-) parity resonances. Inverting these relations and neglecting the small proton-neutron mass difference, we find
\begin{eqnarray}\label{eq:EMratio}
\mbox{spin $\frac{1}{2}$:} &\hspace{1cm}& \frac{\kappa_{ {
N^*}n}}{ \kappa_{ { N^*}p}} = \frac{\mathcal{A}^n_{1/2}}{\mathcal{A}^p_{1/2}} \;, \nonumber\\
\mbox{spin $\frac{3}{2}$:} &\hspace{1cm}& \frac{\kappa_{ {
N^*}n}^{\left( 1 \right)}}{ \kappa_{ { N^*}p}^{\left( 1 \right)}} =
\frac{ \sqrt{3} \mathcal{A}^n_{1/2} \pm \mathcal{A}^n_{3/2}}{\sqrt{3} \mathcal{A}^p_{1/2} \pm
\mathcal{A}^p_{3/2}} \;, \\
& &  \frac{\kappa_{ { N^*}n}^{\left( 2 \right)}}{ \kappa_{ {
N^*}p}^{\left( 2 \right)}} = \frac{ \sqrt{3} \mathcal{A}^n_{1/2} -
\frac{m_p}{m_{ N^*}} \mathcal{A}^n_{3/2}}{\sqrt{3} \mathcal{A}^p_{1/2} -
\frac{m_p}{m_{ N^*}} \mathcal{A}^p_{3/2}} \;.\nonumber
\end{eqnarray}
Note that these conversion rules are only meaningful for $N^{\ast}$'s, since the delta-nucleon magnetic transition moments are isospin independent.

\begin{table}
\caption{Photocoupling helicity amplitudes of selected nucleon resonances in units $10^{-3}\,\text{GeV}^{-1/2}$ from the Bonn relativistic constituent-quark model~\cite{DiplomKreuzer}, the Review of Particle Physics~\cite{pdg} and two SAID analyses (SM95~\cite{SAID96} and SP09~\cite{SAID09}). No experimental information exists for the $P_{13}(1900)$. SP09 provides photo-decay amplitudes to protons and does not find evidence for the $P_{11}(1710)$ resonance~\cite{SAID06}. The ratio of EM couplings to proton and neutron (see Eq.~\eqref{eq:EMratio}) is listed as well. The SP09 ratios are obtained with the $\mathcal{A}^n_J$ of SM95.}
\label{tab:helamp}
\centering
\begin{tabular}{ccrlr@{$\,\pm\,$}rr@{$\,\pm\,$}rr@{$\,\pm\,$}r}
\hline\hline \\ [-2.5ex]
Resonance&&\multicolumn{2}{c}{Bonn} &\multicolumn{2}{c}{RPP} &\multicolumn{2}{c}{SM95} &\multicolumn{2}{c}{SP09}\\\hline \\ [-2.5ex]
$S_{11}(1650)$ &$\mathcal{A}^n_{1/2}$
               &\multicolumn{2}{r}{$-16.00$} &$-15.00$ &$21.00$ &$-15.00$ &$5.00$ &\multicolumn{2}{c}{$-$}\\
               &$\mathcal{A}^p_{1/2}$
               &\multicolumn{2}{r}{$4.30$} &$53.00$ &$16.00$ &$69.00$ &$5.00$ &$9.00$ &$9.10$\\
               &$\frac{\kappa_{ {N^*}n}}{ \kappa_{ { N^*}p}}$
               &\multicolumn{2}{r}{$-3.72$} &$-0.28$ &$0.41$ &$-0.22$ &$0.07$ &$-1.67$ &$1.77$\\\\ [-2.5ex]
$P_{11}(1710)$ &$\mathcal{A}^n_{1/2}$
               &\multicolumn{2}{r}{$-26.70$} &$-2.00$ &$14.00$ &$-2.00$ &$15.00$ &\multicolumn{2}{c}{$-$}\\
               &$\mathcal{A}^p_{1/2}$
               &\multicolumn{2}{r}{$52.80$} &$9.00$ &$22.00$ &$7.00$ &$15.00$ &\multicolumn{2}{c}{$-$}\\
               &$\frac{\kappa_{ {N^*}n}}{ \kappa_{ { N^*}p}}$
               &\multicolumn{2}{r}{$-0.51$} &$-0.22$ &$1.65$ &$-0.29$ &$2.23$ &\multicolumn{2}{c}{$-$}\\\\ [-2.5ex]
$P_{13}(1720)$ &$\mathcal{A}^n_{1/2}$
               &\multicolumn{2}{r}{$-30.20$} &$1.00$ &$15.00$ &$7.00$ &$15.00$ &\multicolumn{2}{c}{$-$}\\
               &$\mathcal{A}^p_{1/2}$
               &\multicolumn{2}{r}{$75.90$} &$18.00$ &$30.00$ &$-15.00$ &$15.00$ &$90.50$ &$3.30$\\
               &$\mathcal{A}^n_{3/2}$
               &\multicolumn{2}{r}{$11.40$} &$-29.00$ &$61.00$ &$-5.00$ &$25.00$ &\multicolumn{2}{c}{$-$}\\
               &$\mathcal{A}^p_{3/2}$
               &\multicolumn{2}{r}{$-25.40$} &$-19.00$ &$20.00$ &$7.00$ &$10.00$ &$-36.00$ &$3.90$\\
               &$\frac{\kappa_{ {N^*}n}^{\left( 1 \right)}}{ \kappa_{ { N^*}p}^{\left( 1 \right)}}$
               &\multicolumn{2}{r}{$-0.39$} &$-2.24$ &$11.60$ &$-0.38$ &$2.00$ &$0.06$ &$0.30$\\
               &$\frac{\kappa_{ { N^*}n}^{\left( 2 \right)}}{ \kappa_{ {N^*}p}^{\left( 2 \right)}}$
               &\multicolumn{2}{r}{$-0.40$} &$0.42$ &$1.15$ &$-0.50$ &$1.08$ &$0.08$ &$0.17$\\\\ [-2.5ex]
$P_{13}(1900)$ &$\mathcal{A}^n_{1/2}$
               &\multicolumn{2}{r}{$2.6$} &\multicolumn{2}{c}{$-$} &\multicolumn{2}{c}{$-$} &\multicolumn{2}{c}{$-$}\\
               &$\mathcal{A}^p_{1/2}$
               &\multicolumn{2}{r}{$5.5$} &\multicolumn{2}{c}{$-$} &\multicolumn{2}{c}{$-$} &\multicolumn{2}{c}{$-$}\\
               &$\mathcal{A}^n_{3/2}$
               &\multicolumn{2}{r}{$16.9$} &\multicolumn{2}{c}{$-$} &\multicolumn{2}{c}{$-$} &\multicolumn{2}{c}{$-$}\\
               &$\mathcal{A}^p_{3/2}$
               &\multicolumn{2}{r}{$2.2$} &\multicolumn{2}{c}{$-$} &\multicolumn{2}{c}{$-$} &\multicolumn{2}{c}{$-$}\\
               &$\frac{\kappa_{ {N^*}n}^{\left( 1 \right)}}{ \kappa_{ { N^*}p}^{\left( 1 \right)}}$
               &\multicolumn{2}{r}{$1.83$} &\multicolumn{2}{c}{$-$} &\multicolumn{2}{c}{$-$} &\multicolumn{2}{c}{$-$}\\
               &$\frac{\kappa_{ { N^*}n}^{\left( 2 \right)}}{ \kappa_{ {N^*}p}^{\left( 2 \right)}}$
               &\multicolumn{2}{r}{$-0.46$} &\multicolumn{2}{c}{$-$} &\multicolumn{2}{c}{$-$} &\multicolumn{2}{c}{$-$}\\
\hline\hline
\end{tabular}
\end{table}

Values for the helicity amplitudes of $S_{11}(1650)$, $P_{11}(1710)$, $P_{13}(1720)$ and $P_{13}(1900)$ are presented in table~\ref{tab:helamp}. The listed numbers are from the RPP~\cite{pdg} and two SAID analyses~\cite{SAID96,SAID09}. It is clear that the photon couplings of those resonances pertinent to our calculations are poorly determined. The extracted values are often incompatible, even after taking into account the considerable error bars. No experimental information is available for the $P_{13}(1900)$. Table~\ref{tab:helamp} also features photon couplings as calculated in the Bonn constituent-quark model~\cite{DiplomKreuzer}. The theoretical predictions for the transition moments of the $S_{11}(1650)$ to neutron (proton) agree favourably with the SAID analysis SM95~\cite{SAID96} (SP09~\cite{SAID09}). When confronting the Bonn model calculations for the $P_{11}(1710)$ and $P_{13}(1720)$ resonances with the SM95 SAID analysis, one notices that the transition moments to proton and neutron are overestimated, while their ratio matches within the error. The Bonn constituent-quark model provides a fair account of all $\mathcal{A}^p_J$ from the SP09 analysis. This analysis, however, finds no evidence for the $P_{11}(1710)$ resonance~\cite{SAID06}.

A crucial constraint for the kaon production amplitude is gauge invariance. It is well-known that the $t$-channel Born diagram by itself does not conserve electric charge. In Ref.~\cite{GuidalPhotoProdKandPi}, an elegant recipe to correct for this was outlined. Adding the electric part of a Reggeized $s$-channel Born diagram ensures that the $p(\gamma,K^+)\Sigma^0$ amplitude is gauge invariant. For the $n(\gamma,K^+)\Sigma^-$ reaction, on the other hand, a gauge-invariant amplitude is obtained by including the electric part of a Reggeized $u$-channel Born diagram.

\begin{figure}
 \centering
 \includegraphics[width=\textwidth]{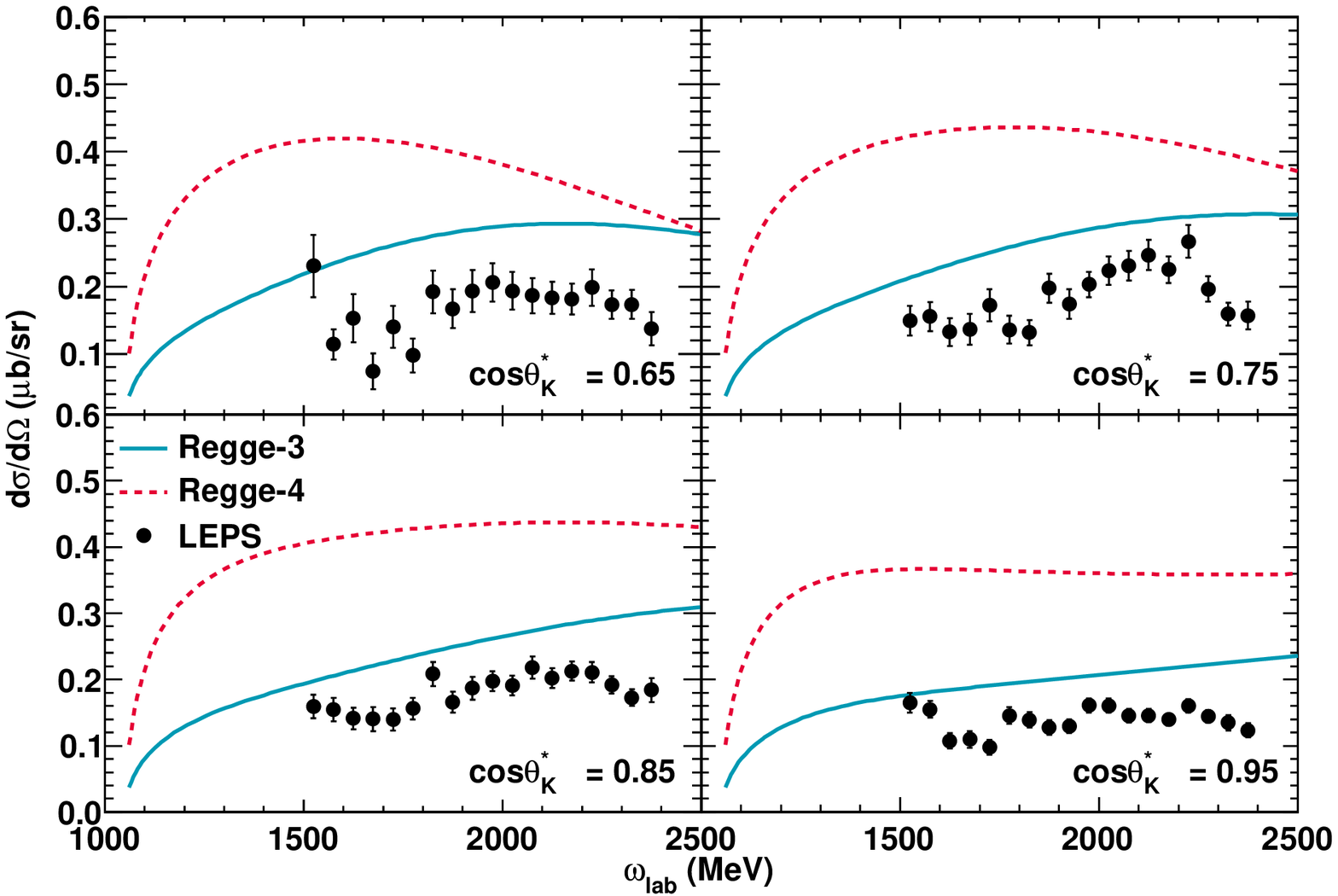}
  \caption{Regge-model predictions for the $n(\gamma,K^+)\Sigma^-$ differential cross section as a function of the incoming photon's lab energy for four different values of the kaon centre-of-mass scattering angle. Data from Ref.~\cite{LEPSiso6}. The error bars represent the statistical uncertainties only. The systematic uncertainty is of the order of $20\,\%$.}
 \label{fig:iso6dcs}
\end{figure}

\begin{figure}
 \centering
 \includegraphics[width=\textwidth]{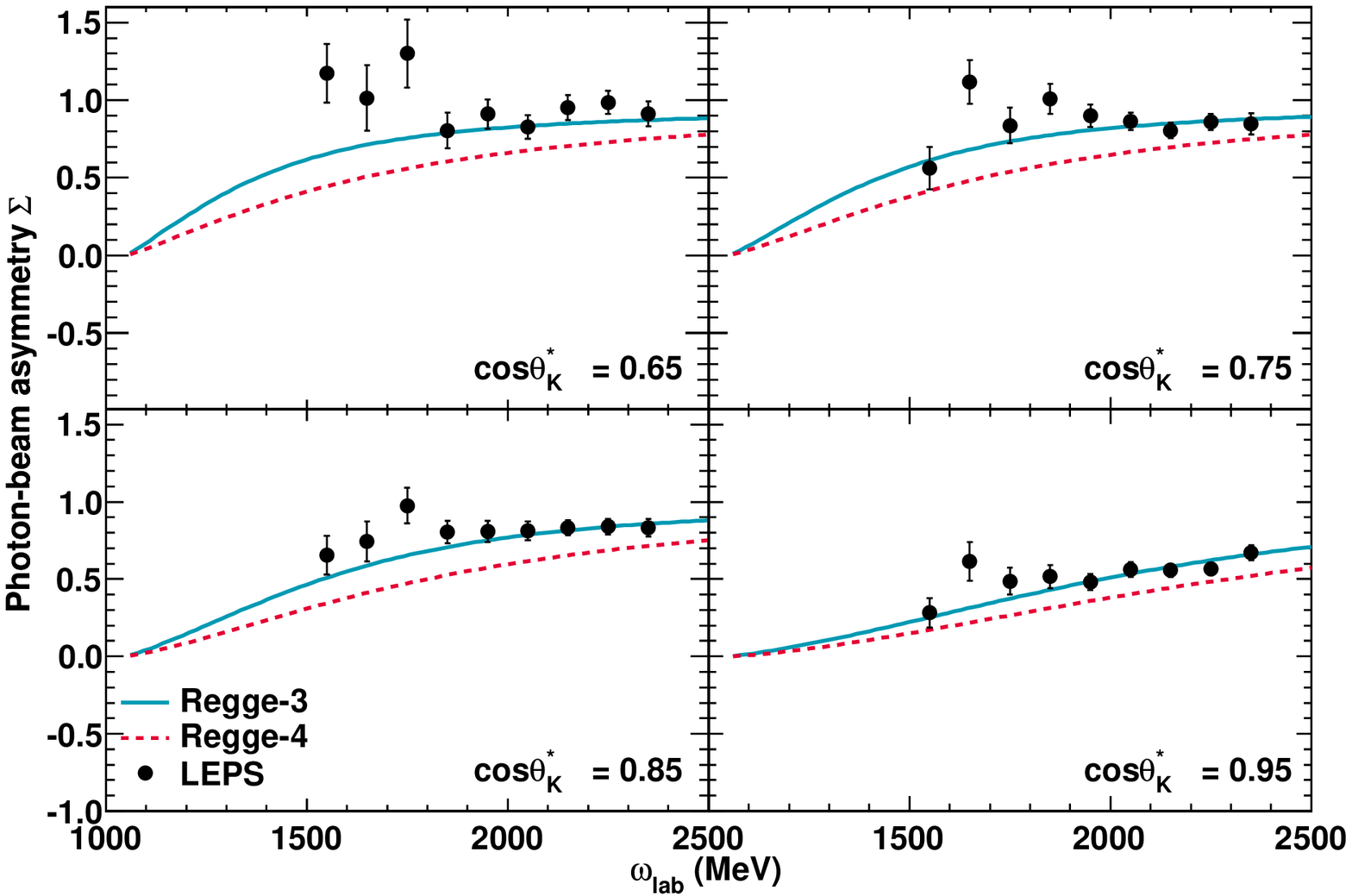}
  \caption{Regge-model predictions for the $n(\gamma,K^+)\Sigma^-$ photon-beam asymmetry as a function of the incoming photon's lab energy for four different values of the kaon centre-of-mass scattering angle. Data from Ref.~\cite{LEPSiso6}. The error bars represent the statistical uncertainties only. The systematics are estimated to be $|\Delta\Sigma|\approx0.2$.}
 \label{fig:iso6pho}
\end{figure}

To our knowledge, the only data for the $n(\gamma,K^+)\Sigma^-$ reaction channel has been published by the LEPS collaboration~\cite{LEPSiso6}. These data comprise differential cross sections and photon-beam asymmetries at forward angles ($\cos\theta_{K}^{\ast}\geq 0.65$) in the energy range $1.5\,\text{GeV}\leq\omega_{\text{lab}}\leq2.4\,\text{GeV}$. This dataset has been obtained through quasi-free kaon photoproduction from a deuterium target. Systematic errors originate from final-state interactions, the pion-mediated two-step process and detector uncertainties. Quadratically summing the estimates given in Ref.~\cite{LEPSiso6} yields uncertainties of the order of $20\,\%$ for the differential cross section and $|\Delta\Sigma|\approx0.2$ for the photon-beam asymmetry. Figures \ref{fig:iso6dcs} and \ref{fig:iso6pho} show our Regge-model predictions for the measured observables. The differential cross section is fairly energy-independent and settles between $0.1$ and $0.2\,\mu\text{b}$. The predictions of the Regge models provide an acceptable description of the data. Overall, the quality of agreement is better for the Regge-3 variant. The Regge-4 model overpredicts the cross section by a factor of two, roughly. The LEPS data shows a clear predilection for the Regge-3 model. In previous work~\cite{RPRsigma,RPRelectro}, we had not been able to discriminate between the two Regge models on the basis of the $p(\gamma^{(\ast)},K^+)\Sigma^0$ data. As can be appreciated from figure~\ref{fig:iso6pho}, both Regge models provide a satisfactory account of the photon-beam asymmetry, with a vanishing asymmetry at threshold and a steady rise as the energy increases. Again, it should be stressed that these results do not involve any free parameters and are anchored to the fitted $p(\gamma,K^+)\Sigma^0$ amplitude through $SU(2)$ isospin symmetry at the strong interaction vertex. Despite its simplicity, our approach can account quantitatively for the LEPS measurements.

Both the differential cross section and the photon-beam asymmetry in figures~\ref{fig:iso6dcs} and~\ref{fig:iso6pho} exhibit a rather smooth energy dependence. Nevertheless, some structure can be discerned in the differential cross section, which can be attributed to nucleon and delta resonances. Their role in the $n(\gamma,K^+)\Sigma^-$ reaction can be evaluated with the RPR amplitude. As was outlined previously, the transformation of the $p(\gamma,K^+)\Sigma^0$ amplitude requires a set of helicity amplitudes. The values extracted in the latest SAID analysis, SP09~\cite{SAID09}, are ill-suited for our purposes, as this analysis does not provide resonance couplings to neutrons. We performed calculations with the two other sets (RPP and SM95), and found them to produce qualitatively similar results. In what follows, we will discuss the representative results obtained with the helicity amplitudes extracted in the SAID SM95 analysis~\cite{SAID96}. No experimental information is available for the $P_{13}(1900)$ resonance. Therefore, we allow the ratios of its magnetic transition moments $\kappa_{ { N^*}n}^{\left( 1,2 \right)}/ \kappa_{ {N^*}p}^{\left( 1,2 \right)}$ (see Eq.~\eqref{eq:EMratio}) to vary between $-2$ and $+2$. This range encompasses the Bonn model predictions. Since the EM transition strengths for delta resonances to protons and neutrons are identical, we include the $D_{33}(1700)$, $S_{31}(1900)$, $P_{31}(1910)$ and $P_{33}(1920)$ resonances with the same EM coupling constants as determined in the $p(\gamma,K^+)\Sigma^0$ reaction channel.

\begin{figure}
 \centering
 \includegraphics[width=\textwidth]{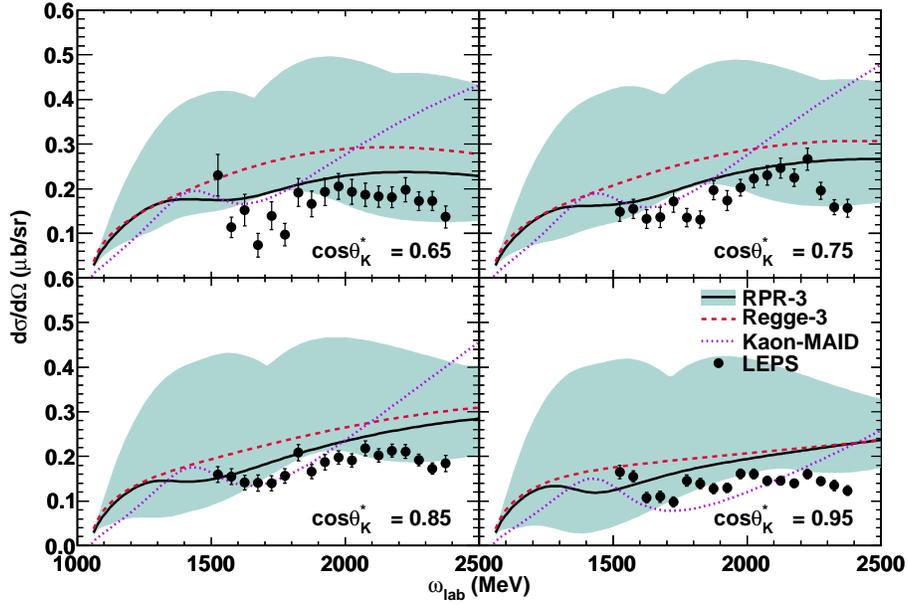}
  \caption{The $n(\gamma,K^+)\Sigma^-$ differential cross section as a function of the incoming photon's lab energy for four different values of the kaon centre-of-mass scattering angle. The dashed curve indicates the Regge-3 model, whereas the full curve corresponds to the RPR-3 amplitude, i.e.\ Regge-3 supplemented with $S_{11}(1650)$, $D_{33}(1700)$, $P_{11}(1710)$, $P_{13}(1720)$, $P_{13}(1900)$, $S_{31}(1900)$, $P_{31}(1910)$ and $P_{33}(1920)$ resonances. The shaded area takes the uncertainties of the adopted helicity amplitudes into account. These uncertainties are listed in table~\ref{tab:helamp} under SM95. The ratios of EM coupling constants for the $P_{13}(1900)$ resonance are taken in the range $[{-2},2]$. The dotted curve represents the Kaon-MAID~\cite{KaonMAID} predictions. Data from Ref.~\cite{LEPSiso6}.}
 \label{fig:iso6rpr_dcs}
\end{figure}

\begin{figure}
 \centering
 \includegraphics[width=\textwidth]{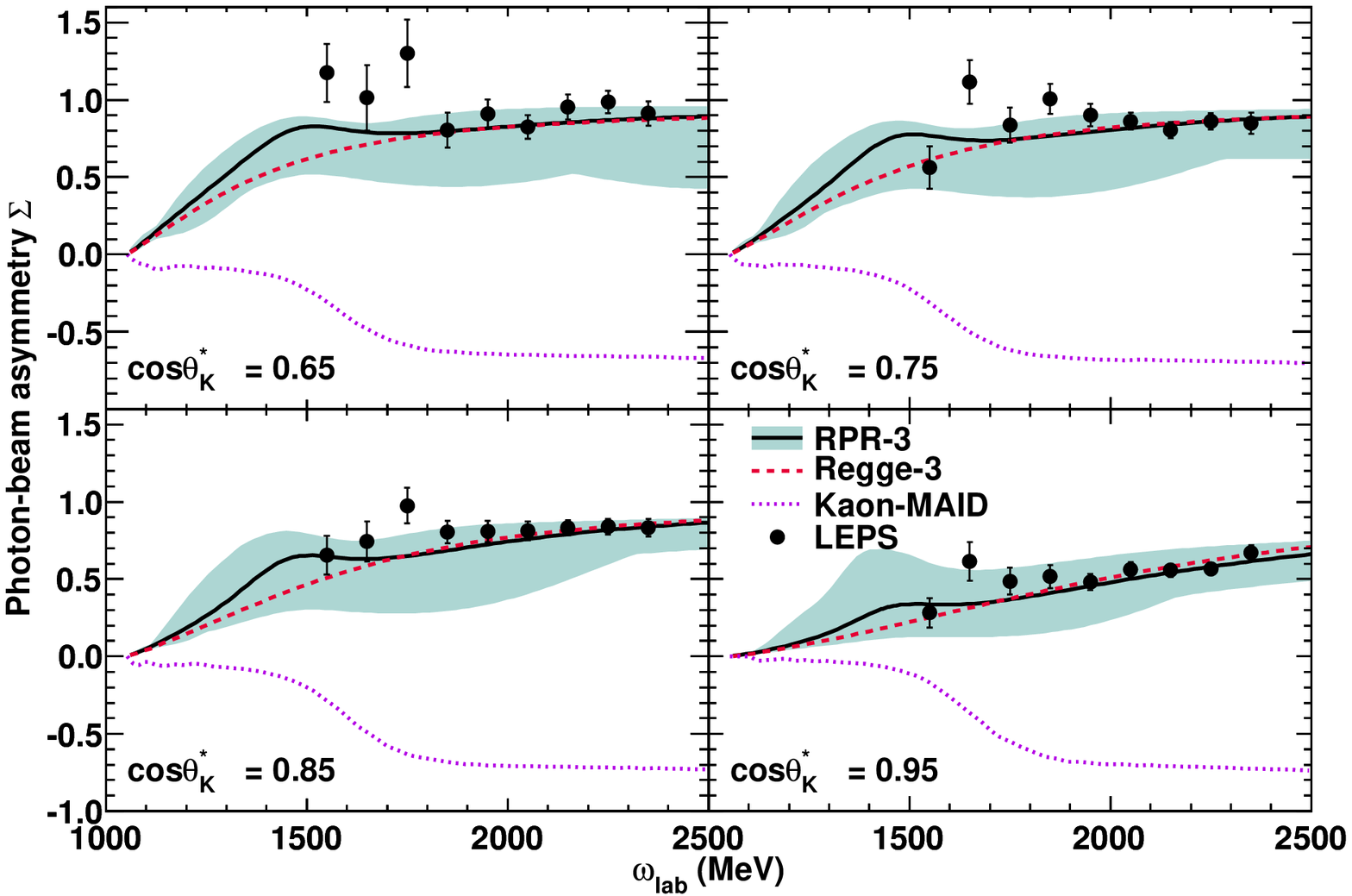}
  \caption{The $n(\gamma,K^+)\Sigma^-$ photon-beam asymmetry as a function of the incoming photon's lab energy for four different values of the kaon centre-of-mass scattering angle. The curves are as indicated in figure~\ref{fig:iso6rpr_dcs}. Data from Ref.~\cite{LEPSiso6}.}
 \label{fig:iso6rpr_pho}
\end{figure}

The amplitudes of the RPR-3 model are the sum of the Regge-3 background and resonance contributions. In figure~\ref{fig:iso6rpr_dcs} and~\ref{fig:iso6rpr_pho}, we confront the RPR-3 and Regge-3 predictions for $n(\gamma,K^+)\Sigma^-$ with the LEPS data. One observes a destructive interference between the Reggeized background and the resonance diagrams. This reduces the reaction strength and marginally improves the agreement with the cross section data in all angular bins and for all energies. From figure~\ref{fig:iso6rpr_pho}, it is plain that the Regge-3 and RPR-3 models provide similar predictions for the photon-beam asymmetry $\Sigma$. This observation leads us to conclude that $\Sigma$ is less sensitive to resonant contributions.

From table~\ref{tab:helamp}, we learn that the helicity amplitudes extracted in the SM95 analysis have considerable error bars. Their impact is assessed in figures~\ref{fig:iso6rpr_dcs} and~\ref{fig:iso6rpr_pho}, and is quite dramatic. Using the error bars given in table~\ref{tab:helamp}, we considered 21 equidistant values for each $\frac{\kappa_{N^{\ast}n}}{\kappa_{N^{\ast}p}}$ of the nucleon resonances in the RPR model. We computed the differential cross sections and photon-beam asymmetries for each of these $21^6$ combinations. The shaded area of figures~\ref{fig:iso6rpr_dcs} and~\ref{fig:iso6rpr_pho} indicates the range of values for $d\sigma/d\Omega$ and $\Sigma$ obtained with this procedure. The experimental ambiguities of the transformed photon couplings result in deviations up to $100\,\%$ for the differential cross section. The photon-beam asymmetry, by comparison, is affected to a smaller extent. Nevertheless, the uncertainty can be as large as $\Delta\Sigma\approx0.8$. 

Regge models turned out to have considerable predictive power, because they can rely solely on isospin arguments when transforming the $K^+$ production amplitude from proton to neutron targets. The RPR model, on the other hand, is less resilient. It is clear that the current errors on the extracted helicity amplitudes impose severe constraints on the predictive power of the RPR model. This result is not limited to the RPR model, but is inherent to any model that includes the exchange of nucleon resonances in the $s$-channel. To illustrate this, we have included model predictions for $n(\gamma,K^+)\Sigma^-$ from Kaon-MAID~\cite{KaonMAID,KMaidpaper} in figures~\ref{fig:iso6rpr_dcs} and~\ref{fig:iso6rpr_pho}. This isobar model treats the background in terms of $s$-, $t$- and $u$-channel Born diagrams as well as $K^{\ast}(892)$ and $K_1(1270)$ exchange. In addition, Kaon-MAID considers a ``minimal'' set of resonances, consisting of $S_{11}(1650)$, $P_{11}(1710)$, $P_{13}(1720)$, $S_{31}(1900)$ and $P_{31}(1910)$. All of these resonances have a 3- or 4-star rating. In order to convert the $p(\gamma,K^+)\Sigma^0$ to the $n(\gamma,K^+)\Sigma^-$ amplitude, SM95 values for the helicity amplitudes were adopted. As can be appreciated from figure~\ref{fig:iso6rpr_dcs}, Kaon-MAID describes the measured differential cross section up to $\omega_{\text{lab}}\approx2000\,\text{MeV}$ at forward angles. The predicted rise of the differential cross section with increasing $\omega_{\text{lab}}$ is absent in the data. Furthermore, the predicted sign of the photon-beam asymmetry is opposite to the data.

In summary, we have presented a Regge-plus-resonance (RPR) approach to $K^+\Sigma^-$ production from the neutron. We model the troublesome background contributions through the exchange of $K^+(494)$ and $K^{\ast+}(892)$ Regge-trajectories. This Regge model can be supplemented with a selection of $s$-channel resonances. In order to gauge the predictive power of kaon production models whose parameters are constrained by data obtained off proton targets, we have confronted RPR and Kaon-MAID predictions with recent $n(\gamma,K^+)\Sigma^-$ data. The conversion to neutron targets of kaon production models that include resonant diagrams requires knowledge of helicity amplitudes. Beyond the second resonance region, the latter are either unknown or poorly constrained by pion production data. As a consequence, they put severe limits on the predictive power of both the RPR and isobar approaches. The Regge model, by contrast, offers an elegant and economical description of electromagnetic kaon production. Isospin symmetry suffices to anchor the neutron to the proton channel. As a result, the Regge model yields more reliable predictions in the $n(\gamma,K^+)\Sigma^-$ channel.

\section*{Acknowledgements} 
This work was supported by the Research Foundation -- Flanders (FWO) and the research council of Ghent University.

\bibliographystyle{elsarticle-num}
\biboptions{sort&compress}
\bibliography{bibliography}

\end{document}